# Applying the deconvolution approach in order to enhance RRDE time resolution: practical issues and limitations.

Artem V. Sergeev[1], Daniil M. Itkis[2], Alexander V. Chertovich[1*]

[1] Physics Department, Lomonosov Moscow State University, Moscow, 119991 Russia

[2] Chemistry Department, Lomonosov Moscow State University, Moscow, 119991 Russia

* Corresponding author. Physics Department, Lomonosov Moscow State University, Leninskie Gory, Moscow, 119991 Russia. Tel.: +7 495 939 4013. E-mail address: chertov@polly.phys.msu.ru

## Abstract

A ring electrode of an RRDE setup is often used to detect a redox active specie produced at the disk electrode. It is especially useful when some side processes occur at the disk (e.g. passivation film growth) along with the main electrochemical reaction that produces the soluble redox-active specie. Unfortunately, the detected ring signal is delayed and smeared so that fast changes of the disk processes intensities cannot be studied. The deconvolution approach is a mathematical data processing procedure that enables reconstruction of the disk signal with hypothetically infinite accuracy. Although there are practical limitations arising mainly from impossibility of exact measurement of the impulse response function and inevitable presence of a noise component in the ring signal used for the reconstruction. In this work a series of calculations were performed in order to investigate the applicability and reliability of the deconvolution approach. Also a procedure to filter out spurious artifacts from the reconstructed disk signal was suggested and tested.

## Introduction

The rotating ring-disk electrode allows one to detect a product of the electrochemical reaction occurring at the disk by facilitating reverse electrochemical reaction at the ring. It is especially useful when more than one processes occur at the disk. The disc current can be divided into two parts: one (the main) that associated with the redox-active product emitted into the solution and that associated with other side processes such as passivation film growth or producing a specie that cannot be detected at the ring. In a steady state regime, the disk current fraction that associated with the main process can be easily determined through simple dividing the ring current by the RRDE collection efficiency. The result can be extracted from the total disk current to yield the side processes fraction. However, if time dependence of the current fractions were of interest, one would have to deal with the limited time resolution of the RRDE. The limitation arises from the fact that the disk signal in form of the redox-active specie emitted into the solution delays and smears before reaching the ring electrode. Thus, fast changes of the disk processes intensities cannot be detected.

The deconvolution approach to the RRDE experiment data analysis allows one to reconstruct the main fraction of the disk current (associated with emission of the redox-active product) as a function of time with hypothetically unlimited time resolution. It is based on the essential properties of the Fourier transform and relies on the impulse response function (IRF) of the RRDE, which can be both measured or calculated.

Measuring a fast changing disk current component is an actual problem for the Li-air battery research. At the Li-air cathode two electrochemical process seem to occur simultaneously: 1st - oxygen reduction to lithium superoxide $LiO_2$ ($O_2 + e^- + Li^+ \rightarrow LiO_2$)[1–4] and 2nd - $LiO_2$ reduction to $Li_2O_2$ ($LiO_2 + e^- + Li^+ \rightarrow Li_2O_2 \downarrow$)[5–7]. The first reduction produces the flux of soluble $LiO_2$ that can be detected by electrooxidation at the ring. Unfortunately, the second reduction partially consumes the product of the first electrochemical reaction and results in growth of a passivating $Li_2O_2$ film. That was an obstacle for applying RRDE for studying oxygen reduction reaction in lithium electrolytes[8].

Upon forming the first molecular layers of the film (i.e. upon transition from a bare electrode surface to $Li_2O_2$ surface), the ratio between the two electrochemical processes can shift drastically and the unprocessed ring current does not reflect such quick changes. The second reduction leading to $Li_2O_2$ film formation is believed to be relatively more intense under high current/overpotential condition[7,9,10]. Thus studying the passivation process implies diffusion limited condition at the disk. Considering oxygen solubility, $O_2$ diffusion coefficient and viscosity of solvent commonly used in Li-$O_2$ studies (e.g. $1.67 \cdot 10^{-5}$ cm$^2$/s[1], 2.1 mM[11], 1.77 cSt[12] respectively in case of DMSO) one can calculate diffusion limited current densities according to Levich equation. Those are about 0.7 - 2.1 mA/cm$^2$ for the electrode rotation speed in the range of 200-1600 rpm. Assuming most of $LiO_2$ produced participate in $Li_2O_2$ film growth the estimated time it takes to form a monolayer of $Li_2O_2$ is 0.47-0.17 seconds for the same rotation speed range 200-1600 rpm. (Considering $Li_2O_2$ molar mass is 46 g/mol, density is 2.14 g/cm$^3$ [13] and the monolayer thickness is about 0.77 nm[13].)

Another vast area where the deconvolution approach can be applied is corrosion of metal surface passivation. During anodic process, the generated metal ions can react with the media that results in passivating film growth e.g. a layer of iron oxides and hydroxides passivating the iron surface[14]. To investigate the very onset of the film growth an enhanced time resolution is required. The deconvolution approach can also be useful to account for adsorption/desorption of redox-active species at the disk electrode as the characteristic time of such processes can be about 0.1-0.5 s[15] which is close to RRDE transient time.

The deconvolution approach was first suggested in [16], but unfortunately the paper only presents the basics of the method and lacks the discussion of the crucial practical issues and applicability. Here we consider the problems one would face trying to employ the deconvolution approach including: measuring IRF and correcting it for the actual values of the redox-active specie diffusion coefficient and the solution viscosity; eliminating spurious artifacts from the reconstructed ring current; estimating the noise level of the ring current and determining time resolution limitation caused by the noise.

**Theoretical basis**

The deconvolution approach is based on the use of impulse response function (IRF) of a RRDE apparatus. Let us think of an abstract system that transforms input signal $x(t)$ into the output signal $y(t)$. If the system is linear and time-invariant, then the transformation can be represented as convolution:

$$y(t) = x(t) * h(t) = \int_{-\infty}^{+\infty} h(\tau)x(t-\tau)d\tau \quad (1)$$

where $h(t)$ is the system's IRF, that is system's output in response to Dirac delta function $\delta(t)$ as an input. According to the convolution theorem the above integration (1) can be carried out in frequency domain by simple multiplication of Fourier transforms:

$$\hat{y}(f) = \hat{x}(f)\hat{h}(f) \quad (2)$$

The latter allows us to solve a reverse problem, i.e. to find input $x(t)$ if the output $y(t)$ and IRF $h(t)$ are known:

$$x(t) = F^{-1}\left\{\frac{\hat{y}(f)}{\hat{h}(f)}\right\} \quad (3a)$$

where $F^{-1}$ denotes an inverse Fourier transform. More details can be found in signal processing literature, e.g. [17].

The stated theory is applicable to RRDE as Faradaic currents and diffusion and convection fluxes are linear in respect to redox specie concentration. (Rotation speed, temperature and other external conditions must be kept constant for the time invariance to hold.)

*Typical application case*

Let's start with a general description of the electrochemical system to which the deconvolution approach is supposed to be applied. Specie A oxidizes at the disk producing a flux $j_d(t)$ of soluble $A^+$ outgoing from the disk into the bulk solution. This process considered to be *main* is accountable for the $I_{d\_main}(t)$ component of the disk current. The reduction of $A^+$ at the ring produces ring current proportional to the flux $j_r(t)$ toward the ring surface.

The *side* reaction results in generation of specie B, that cannot be detected at the ring either because it does not undergo electrochemical reduction at the ring potential or because it stays at the disk surface in the form of a passivation layer. There are as well can be multiple side processes and all their contribution to the disk current is assigned to the $I_{d\_side}(t)$. Fig. 1 illustrates the described system.

DISK: main reaction: $A \rightarrow A^+ + e^-$; side reaction: $A \rightarrow B + e^-$;

$I_d(t) = I_{d\_main}(t) + I_{d\_side}(t)$, $I_{d\_main}(t) = j_d(t)F$, where $F$ is the Farday constant.

RING: $A^+ + e^- \rightarrow A$; $I_r(t) = j_r(t)F$

When applying the deconvolution approach $I_{d\_main}(t)$ (or $j_d(t)$) should be considered as an input signal $x(t)$, while $I_r(t)$ (or $j_r(t)$) – as an output signal $y(t)$. Thus disk main reaction current can be reconstructed as:

$$I_{d\_main}(t) = F^{-1}\left\{\frac{\hat{I}_r(f)}{\hat{h}(f)}\right\} \quad (3b)$$

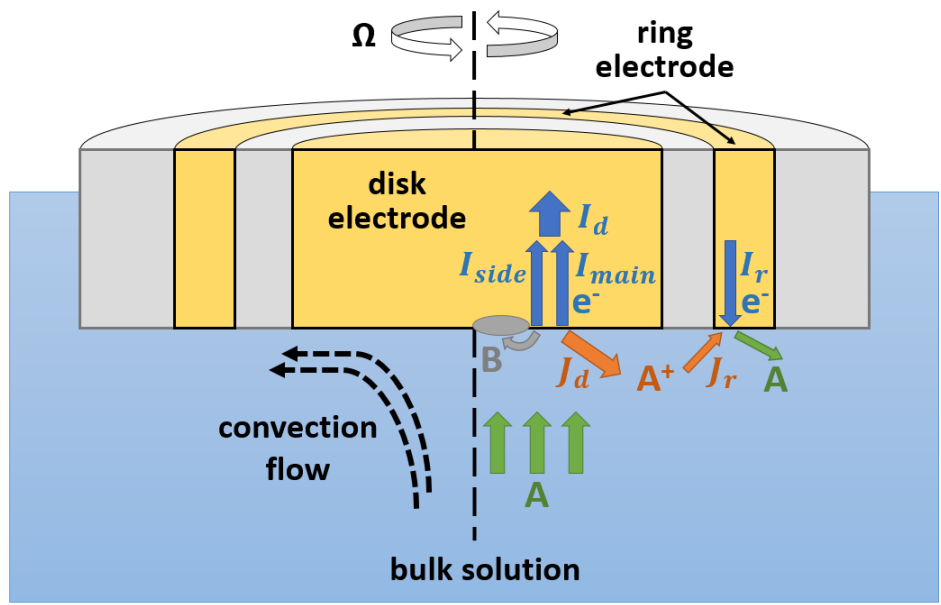

Figure 1. Schematic representation of the considered electrochemical reactions in the RRDE setup.

**Practical Issues**

*Acquisition of IRF*

The typical RRDE's impulse response functions $h(t)$ calculated with the help of a numerical model (see appendix A) for the different values of the rotation speed are presented in Fig. 2. Unfortunately, the IRF cannot be measured directly for the system of interest as it would be affected by the side process (e.g. passivation film growth) that is to be studied. Therefore, the system should be changed to eliminate the side process. That can be done by replacing electrode material, by changing electrolyte composition or replacing the redox-active specie, or by simply measuring IRF at the disk potential values, such that the side process intensity is negligible.

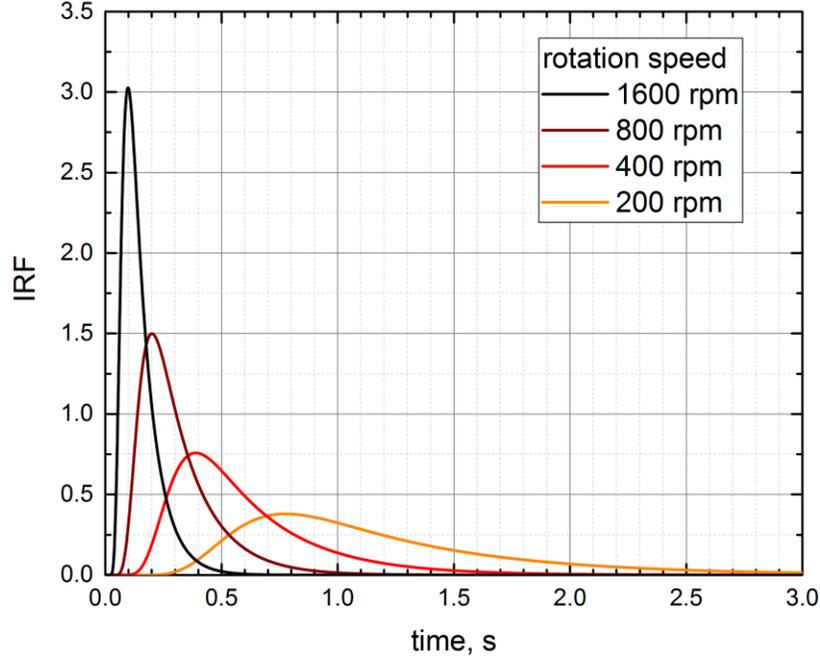

Figure 2. Typical impulse response functions of an RRDE filled with water based electrolyte calculated for the different values of the rotation speed.

If the solution composition is changed, the measured IRF should be corrected for the difference in kinematic viscosity $v$ and diffusion coefficient $D$ (of the redox specie). Luckily, as shown in [18], RRDE transients for different values of $\Omega$, $v$ and $D$ match each other when plotted against dimensionless time $t\omega(D/v)^{1/3}$. IRF is essentially a time derivative of the transient, therefore, upon changing parameters ($\Omega$, $v$ or $D$) it should be rescaled along both axis: abscissa (time) and ordinate (amplitude). Thus, if $h_1(t)$ is measured for a solution with $v_1$ and $D_1$ then IRF for a solution with $v_2$ and $D_2$ can be calculated as:

$$h_2(t) = h_1\left(t \cdot \sqrt[3]{\frac{v_1 D_2}{v_2 D_1}}\right) \cdot \sqrt[3]{\frac{v_1 D_2}{v_2 D_1}} \quad (4)$$

The cubic root dependence decrease sensitivity of the rescaling procedure toward the error of $v$ and $D$ values. The impact of such error on the accuracy of disk flux reconstruction will be investigated later in the text.

It is impossible generate ideal delta-function signal at the disk to measure the IRF. Therefore, one should try to generate as short current pulse as possible and use the following equation to find the IRF:

$$h(t) = \left|F^{-1}\left\{\frac{\hat{J}_r(f)}{\hat{J}_d(f)}\right\}\right| \quad (5)$$

Although, IRF must be real, due to unavoidable addition of noise to the measured $J_d(t)$ and $J_r(t)$ the result of inverse Fourier transform most probably will be a complex signal. Therefore, we recommend to use absolute value.

IRF also can be calculated by means of computer simulation. Original contribution to RRDE numerical simulation was made by Prater and Bard[19]. Since then a lot of efforts were made to develop more sophisticated and precise numerical models that, for example, take into account finite cell dimensions. One can read more on the progress in this area in [20]. Nevertheless, even simple 2-D models are able to predict collection efficiency within 1% accuracy. And off course robustness of all models depends on the accuracy of the input parameters such as redox specie diffusion coefficient (pretty much as IRF rescaling procedure described above).

*Noise influence*

RRDE is essentially a low-pass filter, i.e. high-frequency components of input disk signal $j_d(t)$ are severely suppressed in output ring signal $j_r(t)$. This property also manifests in exponential decay of IRF's Fourier transform $\hat{h}(f)$. The higher the frequency, the smaller the output signal amplitude. In experiments noise inevitably adds to the output signal. Even if overall noise level is quite low, at the high frequencies it overcomes the evanescent true signal. When applying deconvolution approach to reconstruct disk signal the high-frequency components are restored by significant amplification. Mathematically (see eq. 3) it is done by dividing $\hat{y}(f)$ by $\hat{h}(f)$ which tends to zero at high frequencies. In practice high-frequency noise components are amplified to the point where it eclipses the true signal (even strong low-frequency components) and turns the reconstructed disk signal into a complete nonsense. We find this effect to be the major issue of deconvolution approach.

Fortunately, high-frequency noise can be filtered out. The question is what the cut-off frequency $f_{cut}$ should be? Obviously, the higher the $f_{cut}$ the stronger the noise. However, if the cut-off frequency is small, the reconstructed disk signal will lack the sharp (high frequency) details of the original disk signal. As a reference point for the $f_{cut}$ we suggest to use frequency $f_{sn}$, at which ring signal to noise ratio reaches 1.0 (true signal and noise power spectral density values are equal). The $f_{cut}$ should be somewhat smaller than $f_{sn}$ to insure that the noise component of the reconstructed disk signal is significantly weaker than the true signal. A reasonable value is $f_{cut} = f_{sn10}$, where $f_{sn10}$ is the frequency at which true signal spectral density is still 10 times higher than the noise spectral density. The way to estimate $f_{sn}$ is presented in *results and discussion* section, as well as the investigation of $f_{cut}$ influence on the reconstruction results.

*Discreteness and periodicity*

In practice one will have to work with discrete signals and discrete Fourier transforms. Therefore, it is crucial for a successful application of deconvolution approach to bear in mind some features of discrete signal processing. First of all, the sampling interval should be short enough to capture the details of the processes occurring at the disc. $\Delta t$ = 1 ms should be enough as due to the noise presence (as discussed above) the disc signal probably cannot be reconstructed with higher accuracy.

Second, much less obvious note concerns signal periodicity. Let the measured $I_r(t)$ has a finite duration $T$ and is represented by an array containing N = $T/\Delta t$ points. Discrete Fourier transform (which is implemented in almost any scientific software package such as Origin, Matlab or Mathematica) of the signal yields a discrete spectrum $\widehat{I_r}(f)$ containing N points equidistantly placed at interval $\Delta f$ = 1/$T$. The inverse Fourier transform of a discrete spectrum is a periodic signal with period $T = 1/\Delta f$. Thus, original finite data array is actually processed as an infinite periodic signal! Therefore, one should insure that periodically replicated ring signal $I_r(t)$ does not contain any unphysical discontinuity, as it would produce artifacts in the reconstructed disk signal. Such discontinuity arises if $I_r(0) \neq I_r(T)$ (see Fig. 3). Most probably $I_r(0) = 0$. Thus, before finishing the measurement one should put current to zero and wait until ring current also reaches the zero value.

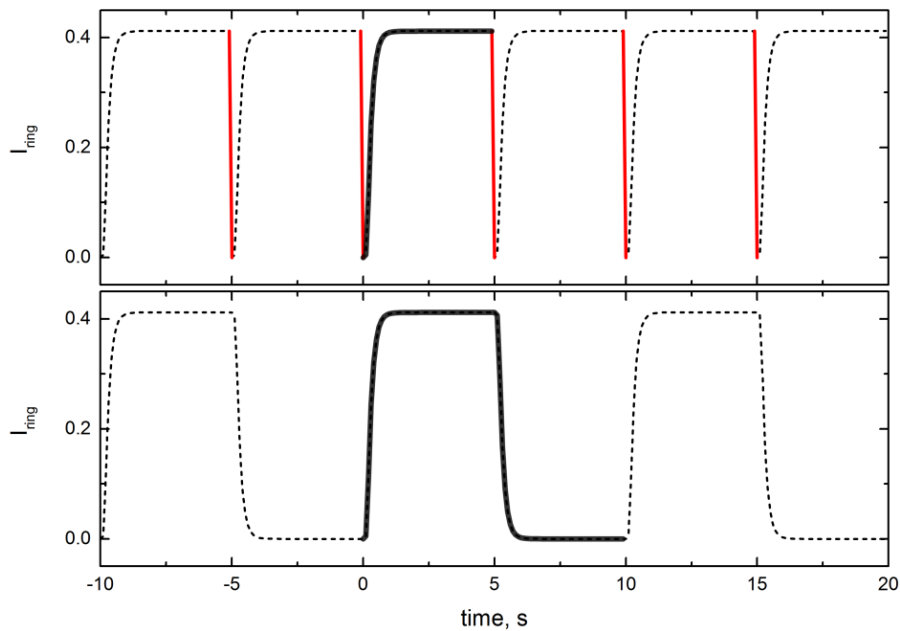

Figure 3. Examples of periodical replication of the measured ring signal. Thick line – original measured signal, thin line – replica. If the signal does not go to zero at the end of the measurement (upper panel) discontinuities arise upon replication (marked red). Otherwise (lower panel) the replicated signal retains its continuity. Disk signal used to calculate the ring response is presented in Fig. 4.

**Methods**

In this work we employed numerical simulation methods to test deconvolution approach and address the above mentioned issues. Ring signals were calculated for a rectangular disk signals by means of convolution with the calculated IRF (eq. 1). To investigate the noise impact on the disk signal reconstruction results a white Gaussian noise model was employed. All signal data sets consisted of 1000 points representing (in time domain) duration $T = 10$ s with sampling interval $dt = 0.01$ s. The Fourier transforms of the signals also consisted of 1000 data points representing frequency span up to 50 Hz (symmetrical spectrum [-50, 50], frequency step $df = 0.1$ Hz).

A 2-D computer model of RRDE system was developed and used to calculate the IRF $h(t)$ (see Fig. 2). The model implements a numerical solution of differential equations describing diffusion and convection phenomena in the framework of the finite differences method. The system of convection-diffusion equations[18] relies on the approximate analytical solution for the flow velocities derived by Karman[21] and Cochran[22]. Spatial grid step size along the radial axis (in the plane of the electrode surface) was 25 μm. In all calculations there were 100 grid nodes along the normal axis (perpendicular to the surface) stretched across distance equal to 1.6 of the diffusion layer thickness $\delta_0$ ($\delta_0 = 1.61 \, D^{1/3} v^{1/6} \omega^{-1/2}$) depending on the specie diffusion coefficient, solution kinematic viscosity and electrode rotation speed. The time step also depended on the system parameters so there were 10000 steps per characteristic time $\tau = 0.51^{-2/3}(v/D)^{1/3}\omega^{-1}$. A ferrocyanide/ferricyanide redox couple in concentrated aqueous electrolyte was chosen as a reference system for the parameters values used in the simulations (i.e. ferrocyanide(ferricyanide) diffusion coefficient $D$ = 0.67(0.72) cm²/s, kinematic viscosity $v$ = 0.9 cSt). RRDE dimensions: disk diameter $d_1$= 4 mm, ring inner diameter $d_2$= 5 mm, ring outer diameter $d_2$= 7 mm. The calculated collection efficiency error falls in less than 1% in respect to a well-recommended theoretical prediction[23].

It should be noted that the same set of calculated IRFs is used here to both produce ring signal (by convolution with the disk signal) and to reconstruct the disk signal. Thus in this paper we investigate not the accuracy of the calculated IRFs but rather the robustness of the deconvolution approach in respect to the error of the known $D/v$ value and the ring noise power.

**Results and Discussion**

As it was explained when discussing IRF acquisition issue, to correct $h(t)$ for the change of solution composition or to calculate $h(t)$ by means of a computer simulation one has to know solution kinematic viscosity $v$ and diffusion coefficient $D$ of the redox active specie. Here we investigate the impact of diffusion coefficient value error on the disk signal reconstruction accuracy. (Here we only talk about $D$ assuming $v$ to be known exactly, though all of the following discussion is true for the ratio $D/v$.)

A broad rectangular pulse of a unit amplitude with period $T = 10$ s and duration $T/2$ was used as an input disk signal $I_d(t)$. A square step signal contains high-frequency components and therefore it is suitable to evaluate reconstruction quality of sharp signal details. Long pulse duration is needed to avoid overlapping of the rise and fall responses. Ring signal $I_r(t)$ was produced by convolution (eq. 1) with reference IRF $h_{ref}(t)$ calculated for a reference diffusion coefficient value $D_{ref}$. Then IRF was corrected (or rather corrupted) according to eq. 4 for a range of diffusion coefficient values. The corrected $h_{cor}(t)$ were used to reconstruct disk signal. An example of reconstructed $I_{d\_rec}(t)$ for $D = 1.5 D_{ref}$ is presented in Fig. 4. Mean square error was used as a measure of reconstruction accuracy. Reconstruction with reference IRF $h_{ref}(t)$ resulted in MSE ~ 10⁻⁸ which is practically ideal. The resulted MSE values are plotted in Fig. 5 against diffusion coefficient values used to correct IRF.

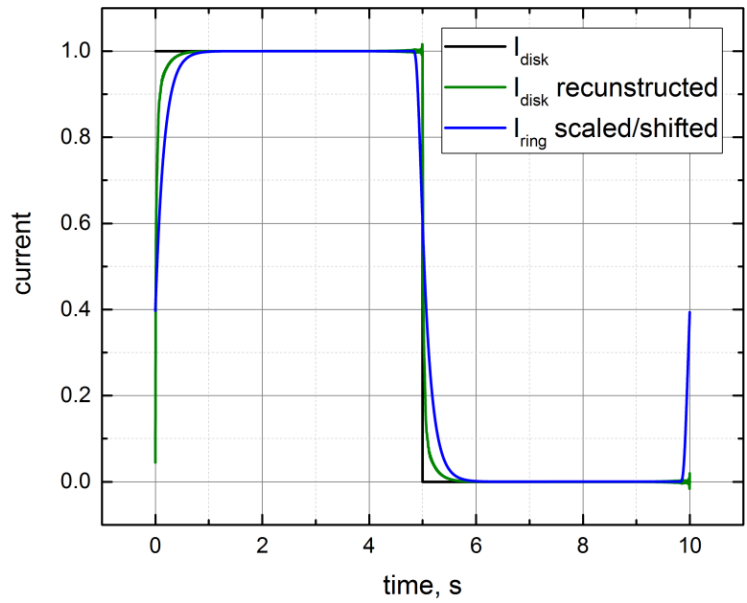

Figure 4. Disk current: true signal (black), signal reconstructed by means of the deconvolution approach (green) and signal estimated by rescaling and shifting the ring signal (blue). An overestimated (x1.5) value of the diffusion coefficient was used to calculate IRF for the reconstruction procedure. Electrode rotation speed – 800 rpm.

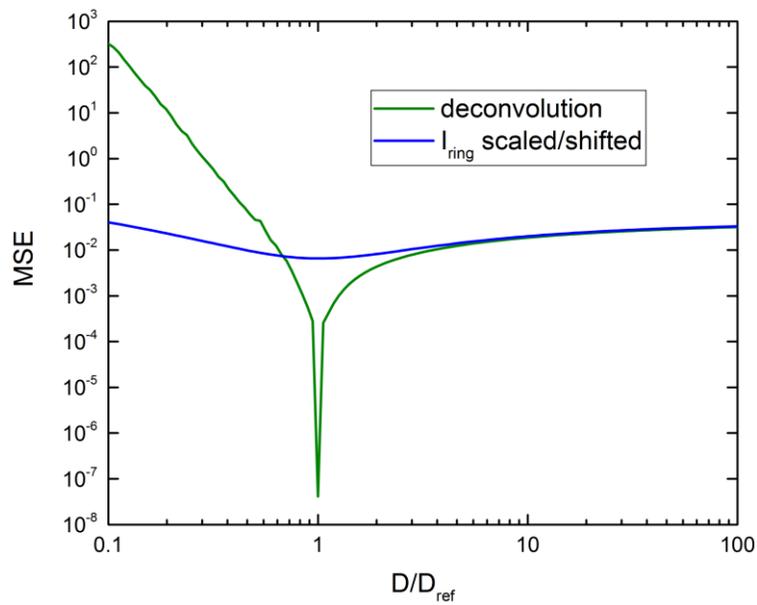

Figure 5. Mean square error of the reconstructed disk signal as a function of the diffusion coefficient value (related to the true value) used for the reconstruction. Green line – reconstruction by means of the deconvolution approach; blue line – estimation by scaling and shifting the ring signal.

To evaluate effectiveness of deconvolution approach we compare it to the ring signal shift approach. Disk signal can be estimated simply by shifting back the delayed ring signal and scaling it in order to allow for the collection efficiency $N_{col}$:

$$j_{d\_shift}(t) = j_r(t - \tau_{shift})/N_{col} \quad (6)$$

where $\tau_{shift}$ is an estimated disk-to-ring delay. A reasonable guess for the $\tau_{shift}$ is the transit time[18]

$$\tau_{transit} = 3.58(v/D)^{1/3}(\ln r_2/r_1)^{2/3}/\Omega \quad (7)$$

According to our calculation the best result in terms of mean square error (MSE) is achieved for $\tau_{shift} = 1.58\tau_{transit}$. (The result of such scale/shift approach is presented in Fig. 4.) The shift time $\tau_{shift}$ also depends on the diffusion coefficient value. The MSE calculated for the ring signal shift approach is presented in Fig. 5 as a function of $D$ value used to estimate $\tau_{shift}$. It should be noted, that the data represented if Fig. 5 for both approaches are independent of rotation speed $\Omega$. The deconvolution approach significantly outperforms the ring signal shift approach if the IRF is accurately corrected for the diffusion coefficient value. Unfortunately, it loses its effectiveness if the estimated $D$ is 1.5 times under-/overestimated. Nevertheless, the acceptable error window is wide enough to apply deconvolution approach in practice.

To investigate noise influence on the reconstruction accuracy we employed white noise model. Noise signal $I_n(t)$ was generated as a random sequence according to standard normal distribution. The ratio between the total energies of the noise and the true ring signal (as in Fig. 3 lower panel) turned out to be $\mathrm{E}_{noise}/\mathrm{E}_{signal} \approx 1$. Than the noise signal $I_n(t)$ was scaled to have certain power. In the example presented in Fig. 6, $I_n(t)$ was scaled so that the $\mathrm{E}_{noise}/\mathrm{E}_{signal} = 10^{-4}$ and added to the ring signal. The resulted noisy signal $I_{r\_n}(t)$ was used to reconstruct disk signal with the help of deconvolution approach. The cutoff frequency $f_{cut} = f_{sn10} = 6.2$ Hz was used to filter out high frequency components of the reconstructed signal $I_{d\_rec}(t)$. Additionally we suggest to smooth $I_{d\_rec}(t)$ by a running average filter with window width $T_{ave} = 1/f_{cut}$. This procedure helps to rid of the artifacts arose from cutting of high-frequency components of the true signal. As one can see in Fig. 5 the resulted $I_{d\_rec}(t)$ much better resembles original $I_d(t)$, than the shifted and scaled ring signal $I_{d\_shift}(t)$ calculated with eq. 6. The reconstructed $I_{d\_rec}(t)$ has about 4 times steeper raise and fall than the ring signal. That supports that deconvolution approach significantly enhances time resolution of RRDE experiments.

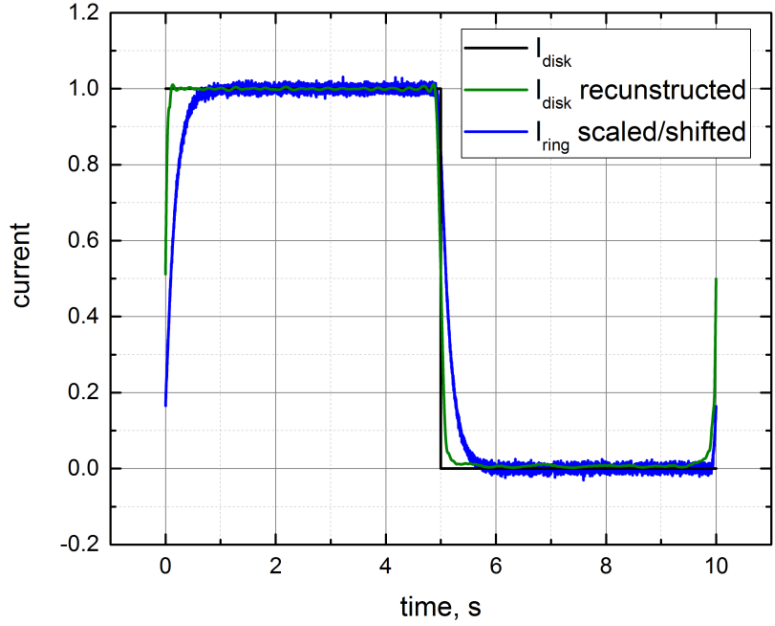

Figure 6. Disk current: true signal (black), signal reconstructed by means of the deconvolution approach (green) and signal estimated by rescaling and shifting the ring signal (blue). Noise was added to the true ring signal before reconstruction. Relative noise energy $E_{noise}/E_{signal} = 10^{-4}$. Electrode rotation speed – 800 rpm.

Minimum characteristic time of a process, that can be studied with the help of deconvolution approach can be estimated as $T_c = 1/f_{cut}$. Thus $T_c$ (as well as $f_{cut}$) depends on the noise spectral density. Adopting a white noise approximation (uniform power spectral density) we calculated $f_{cut}$ as a function of relative noise energy $E_{noise}/E_{signal}$ (see fig. 7). $f_{cut}$ also depends on the RRDE parameters in similar way as way $\tau_{transit}$ (eq. 7), so one can use following relation to recalculate it:

$$f_{cut\_2} = f_{cut\_1}\left(\frac{\tau_{transit\_1}}{\tau_{transit\_2}}\right) \quad (8)$$

Naturally $f_{cut}$ and time resolution ability grow proportionally to rotation speed $\Omega$ and $(D/\nu)^{1/3}$.

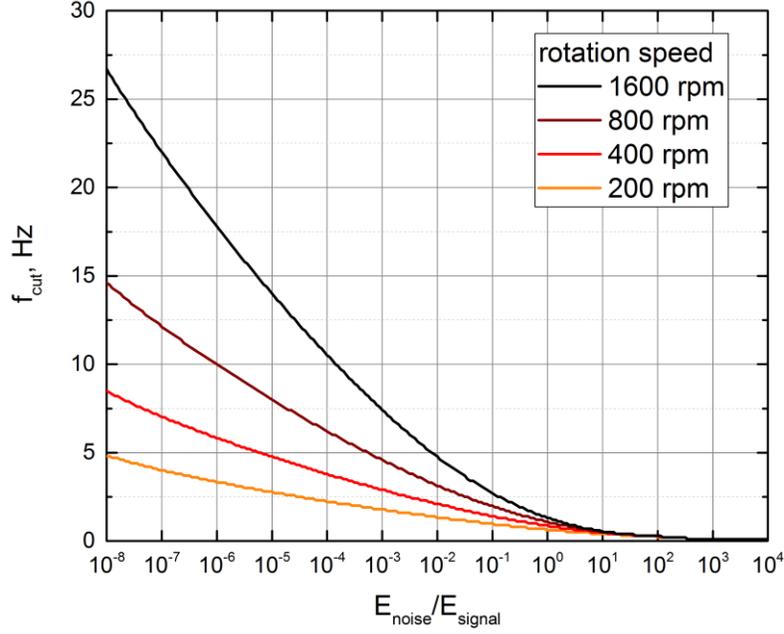

Figure 7. Cut-off frequency as a function of the noise energy related to the true ring signal energy. Electrode rotation speed: 200, 400, 800 and 1600 rpm.

To find $f_{cut}$ one has to know spectral power density of the noise, but experiment only gives us a mix $I_{r\_n}(t)$ of the true ring signal $I_r(t)$ and the noise $I_n(t)$. Luckily, if noise has a random nature, its spectral density can be estimated by averaging $I_{r\_n}(t)$ over several experiments. Upon averaging absolute value of noise Fourier transform should decrease according to:

$$\left|\langle \widehat{I_n}(f) \rangle\right| = \langle |\widehat{I_n}(f)| \rangle / \sqrt{M} \quad (9a)$$

where $M$ is a number of data sets (experiments) to average over. Thus average noise amplitude for a certain frequency $f$ can be estimated as:

$$\langle |\widehat{I_n}(f)| \rangle = \langle |\widehat{I_{r\_n}}(f) - \langle \widehat{I_{r\_n}}(f) \rangle| \rangle \left( \sqrt{M}/(\sqrt{M}-1) \right) \quad (9b).$$

Than the estimated noise spectral density $P_n(f) = \langle |\widehat{I_n}(f)| \rangle^2 / T$ should be plotted along with averaged ring signal spectral density $P_{r\_n}(f) = \left| \langle \widehat{I_{r\_n}}(f) \rangle \right|^2 / T$ to estimate the $f_{cut}$ at which a desired SNR is reached. The example for $M = 4$ based on white Gaussian noise model is presented in Fig. 8. The crossing between the estimated $P_n(f)$ and $P_{r\_n}(f)$ at which SNR ≈ 1 can be clearly seen and located close to the crossing of spectral densities of the true ring signal and the noise. The $f_{cut}$ found with the described technique tends to be underestimated. That reduces deconvolution approach time resolution, though protects one from letting noise components into the reconstructed disk signal.

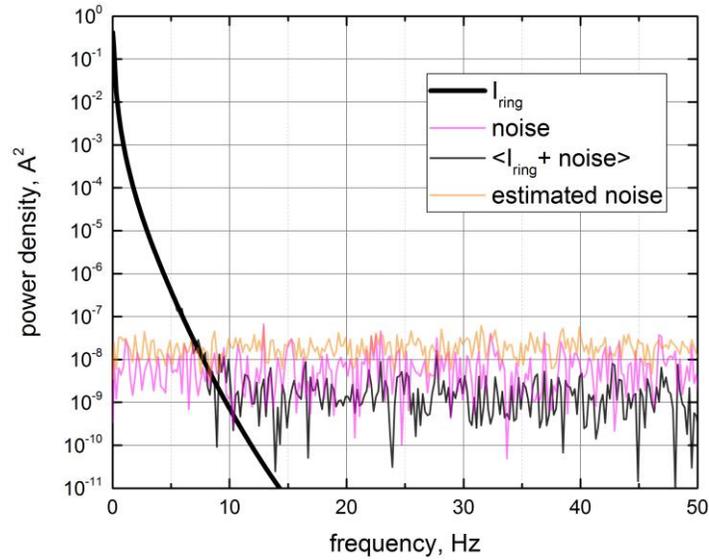

Figure 8. Power spectral density of the true ring signal (bold black), the noised ring signal averaged over 4 realizations (thin black), the noise signal – single realization (magenta) and the estimated noise signal (orange). Relative noise energy $E_{noise}/E_{signal} = 10^{-4}$. Electrode rotation speed – 800 rpm.

To analyze $f_{cut}$ overestimation we repeated the procedure for a range of $M$ values (see Fig. 9). For the small values of $M$ the estimated $f_{cut}$ is not fixed due to the randomness of the noise. So for every value of $M$ the calculations were repeated 100 times to assert the standard deviation of the estimated $f_{cut}$ (shown by error bars in Fig. 9). 8-10 experiment iterations is enough to obtain a good $f_{cut}$ estimation. Nevertheless, even two iteration yield an acceptable result, as underestimation of the $f_{cut}$ is not critical for the deconvolution approach.

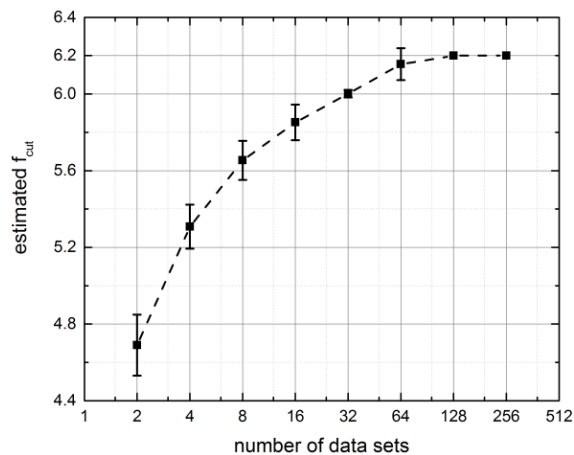

Figure 9. Cut-off frequency calculated based on the estimated noise power spectral density. The estimation performed using certain number of data sets (noise component realizations). The error bars denote standard deviation of the estimated cut-off frequency resulted from the random nature of the noise. Relative noise energy $E_{noise}/E_{signal} = 10^{-4}$. Electrode rotation speed – 800 rpm.

**Conclusions**

In this paper we discussed practical issues of the deconvolution approach application in order to reconstruct disk current fraction associated with redox specie flux outgoing from the disk as a function of time. The reconstruction procedure can be summarized as follows:

1) Calculating IRF according eq. 5. That includes measuring the ring signal in response to a short current pulse at the disk in. The solution composition and/or disk potential should be chosen to eliminate side reactions. The ring potential should be set to guarantee diffusion limited condition.

2) Correcting IRF according to eq. 4 if the solution composition in the system of interest is differ from the solution composition used for measuring IRF.

3) Performing experiment of interest in which side reaction are supposed to take place. Any current/potential signal (including time dependent function) can be applied at the disk. The ring potential should be set to guarantee diffusion limited condition. The ring current should decay to zero in the end of the measurement (i.e. disk current should be set to zero before finishing the measurement) in order to avoid discontinuities upon periodical replication. The experiment should be repeated several times (4-8 is recommended).

4) Estimating the noise power spectral density using eq. 9b ($P_n(f) = \langle |\widehat{I_n}(f)| \rangle^2 / T$) and plotting it along with the powers spectral density of the averaged ring signal ($P_r(f) = |\langle \widehat{I_r}(f) \rangle|^2 / T$) in order to find the cut-off frequency $f_{cut}$ at which $P_n(f)$ is still several times higher (10 is recommended).

5) Reconstructing disc current component $I_{d\_main}(t)$ (associated with generation of the oxidized(reduced) dissolved form of the redox-active specie) according to eq. 3b.

6) Filtering out high frequency components ($f > f_{cut}$) of the reconstructed disc current. That can be done by putting truncating Fourier transform of the signal ($\widehat{I_{d\_main}}(f > f_{cut}) := 0$) and additionally applying a running average filter with the window width $T_{ave} = 1/f_{cut}$.

7) The intensity of the side processes can be found extracting the reconstructed $I_{d\_main}(t)$ from the total disk current: $I_{d\_side}(t) = I_d(t) - I_{d\_main}(t)$.

It was shown that the error of the diffusion coefficient $D$ and viscosity $\nu$ values affects the validity of the IRF corrected for the solution composition change or calculated by means of a computer simulation. According to the calculations deconvolution approach retains its effectiveness if the under-/overestimation of the $D/\nu$ ratio does not exceed 1.5 folds.

Noise influence on the reconstruction result was also analyzed. Filtering out high frequency components was suggested to get rid of spurious artifacts that arise from the noise present in the ring signal. The maximum cutoff frequency which determines the time resolution of the reconstruction procedure was shown to depend on the noise spectral density. Finally, a technique was proposed to estimate the noise spectral density through averaging over several experimental data sets. According to the calculations based on the white Gaussian noise model even 2 experiments is enough to roughly estimate the cutoff frequency.

The analysis and calculations reported in the paper indicate that the deconvolution approach is viable way to significantly enhance RRDE experiments time resolution and has reasonable demands in terms of $D/\nu$ ratio error and noise level.

## Acknowledgements

The work was financially supported by Centre or Electrochemical Energy of Skolkovo Institute of Science and Technology.